\documentstyle[12pt]{article}

\begin{document}

\title{Anomalies and nonperturbative results.}

\author{P.~I.~Pronin
\thanks{E-mail: $petr@theor.phys.msu.su$}
and K.~V.~Stepanyantz
\thanks{E-mail: $stepan@theor.phys.msu.su$}}

\maketitle

\begin{center}
{\em Moscow State University, Physics Faculty,\\
Department of Theoretical Physics.\\
$117234$, Moscow, Russian Federation}
\end{center}

\begin{abstract}
We investigate nonperturbative effects in N=1 and N=2 supersymmetric
theories using a relation between perturbative and exact anomalies as a
starting point. For N=2 supersymmetric $SU(n)$ Yang-Mills theory we
derive the general structure of the Picard-Fuchs equations; for N=1
supersymmetric Yang-Mills theories we find holomorphic part of the
superpotential (with gluino condensate) exactly.
\end{abstract}

\pagebreak
\tableofcontents
\pagebreak


\section{Introduction}
\hspace{\parindent}

The presence of quantum anomalies in the field theory is known for
a long time and plays an impotant role in the high energy physics
\cite{anomalybook}. However, they were usually studied in the frames of
perturbation theory. Only recently the exact expression for R-anomaly was
found in N=2 supersymmetric Yang-Mills theory \cite{matone,howe}. Due to
the instanton contributions it differs from the perturbative result. Such
possibility was pointed rather long ago \cite{shifmanomaly}, but a series
of instanton corrections with unknown coefficients produced considerable
difficulties, in particular, in the research of anomalies cancellation.
So, obtaining of exact results becomes very important. Their derivation
in a number of papers \cite{matone,howe,sonnenschein,eguchi} is based on
the exact results of Seiberg and Witten \cite{seiberg}, but the result
appeared to have a very simple interpretation: exact anomaly is a vacuum
expectation value of the perturbative one. Nevertheless, for checking this
relation one should essentially use the exact expression for the
prepotential, found in \cite{seiberg} by completely different methods.
Thus, we come to the question, whether it is possible to
solve the inverse problem, i.e. to derive exact results from the form of
anomalies. In the present paper we try to do it for supersymmetric
Yang-Mills theories. For N=2 $SU(N_c)$ the presented approach allows to
derive the general structure of Picard-Fuchs equations. The investigation
of N=1 theories turns out to be very similar. In particular, the holomorphic
part of the N=1 superpotential is found to satisfy Picard-Fuchs equation,
that can be solved exactly.

Our paper is organized as follows: Section \ref{susy} is devoted to the
brief review of necessary information concerning N=1 and N=2 supersymmetric
Yang-Mills theories (for details see \cite{shifmanrev,klemm_review}). In the
Section \ref{instanton} we show, that the collective coordinate measure
is not invariant under $U(1)_R$-transformations. Using its transformation
law we are able to define the general structure of nonperturbative
corrections, that agrees with instanton calculations. In the Section
\ref{rel} we derive the relation between perturbative and exact anomalies.
First, in the Subsection \ref{repeat} we reobtain the exact expression for
the R-anomaly in the N=2 supersymmetric SU(2) Yang-Mills theory
\cite{matone,howe} as an indication to the result. Then in the Subsection
\ref{condition} suggested relation is formulated and proven. The Section
\ref{section3} is devoted to consequences. In the Subsection \ref{n2susy}
>from the relation between perturbative and exact anomalies we derive the
general structure of Picard-Fuchs equations and restrictions on their form.
Then the presented approach is applied to N=1 supersymmetric Yang-Mills
theory. In the Subsection \ref{general_structure} we investigate the
holomorphic part of the superpotential and find its structure. The exact
result is obtained in the Subsection \ref{exact_result}. Conclusion is
devoted to the brief review and discussion of the results. Some auxiliary
facts are given in the Appendix.


\section{Supersymmetric Yang-Mills theories}
\label{susy}

\subsection{N=1 supersymmetry}
\label{n1}
\hspace{\parindent}

The massless N=1 supersymmetric Yang-Mills theory with $SU(N_c)$ gauge group
and $N_f$ matter multiplets is described by the action

\begin{equation}\label{N1_action}
S=\frac{1}{16\pi} \mbox{tr\ Im}\left(\tau \int d^4x d^2\theta\ W^2\right)
+\frac{1}{4} \int d^4x d^4\theta \sum\limits_{A=1}^{N_f}
\left(\phi^{+}_A e^{-2V}\phi^A + \tilde\phi^{+A} e^{2V} \tilde \phi_A
\right)
\end{equation}

\noindent
where the matter superfields $\phi$ and $\tilde\phi$ belong to fundamental
and antifundamental representations of the gauge group $SU(N_c)$.

Here we use the following notations

\begin{eqnarray}
&&A_\mu=e A^a_\mu T^a\qquad \mbox{and so on,}\qquad
\mbox{tr} T^a T^b = \delta^{ab};\nonumber\\
&&\tau = \frac{\theta}{2\pi}+\frac{4\pi i}{e^2};
\end{eqnarray}

\begin{eqnarray}\label{notat}
&&V(x,\theta)=-\frac{i}{2} \bar\theta \gamma^\mu\gamma_5 \theta
A_\mu(x) + i \sqrt{2}(\bar\theta \theta)(\bar\theta\gamma_5\lambda(x))
+ \frac{i}{4} (\bar\theta \theta)^2 D;\nonumber\\
&&W(y,\theta)=\frac{1}{2}(1+\gamma_5)\Big(
i\sqrt{2}\lambda(y) + i\theta D(y)
+\frac{1}{2}\Sigma_{\mu\nu}\theta F_{\mu\nu}(y)\nonumber\\
&&\qquad\qquad\qquad\qquad\qquad\qquad\qquad\qquad\qquad
+\frac{1}{\sqrt{2}}\bar\theta(1+\gamma_5)\theta
\gamma^\mu D_\mu \lambda(y)\Big);\nonumber\\
&&\phi(y,\theta)=\varphi(y)+\sqrt{2}\bar\theta(1+\gamma_5)\psi(y)
+\frac{1}{2}\bar\theta_1(1+\gamma_5)\theta f(y);\nonumber\\
&&y^\mu = x^\mu + \frac{i}{2} \bar\theta \gamma^\mu\gamma_5\theta.
\end{eqnarray}

\noindent
Eliminating auxiliary fields we find that in components the action
(\ref{N1_action}) is written as

\begin{eqnarray}
&&S=\frac{1}{e^2} \mbox{Re\ tr} \int d^4x \left(-\frac{1}{4} F_{\mu\nu}
F^{\mu\nu} - i \bar\lambda(1+\gamma_5)\gamma^\mu D_\mu\lambda
+\frac{\theta e^2}{32\pi^2} F_{\mu\nu}\tilde F_{\mu\nu}
\right)
\nonumber\\
&&+\sum\limits_A\int d^4x \left\{
\vphantom{\frac{1}{2}}
D_\mu \varphi^{+}_A D^\mu \varphi^A
+ D_\mu \tilde\varphi^{+}{}^A D^\mu \tilde\varphi_A
+i \bar\Psi \gamma^\mu D_\mu \Psi
\right.
\nonumber\\
&&
\vphantom{\int}
-i \bar\Psi_A (1-\gamma_5) \lambda\varphi^A
+i \varphi^{+}_A \bar\lambda (1-\gamma_5) \Psi^A
-i \tilde\varphi^{+A}\bar\Psi_A (1+\gamma_5) \lambda
+i \bar\lambda (1-\gamma_5) \Psi^A \tilde\varphi_A
\nonumber\\
&&
\left.
+\frac{1}{2} \left(\varphi^{+}_A T^a \varphi^A
- \tilde\varphi^{+A} T^a \tilde\varphi_A \right)^2
\right\}
\end{eqnarray}

\noindent
where we introduced the Dirac spinor

\begin{equation}
\Psi\equiv \frac{1}{2}
\Big[(1+\gamma_5)\psi+(1-\gamma_5)\tilde\psi\Big]
\end{equation}

In the massless case the action is invariant under the transformations

\begin{eqnarray}
&&U(1)_1:\quad
W(\theta) \to e^{i\alpha} W(e^{-i\alpha\gamma_5}\theta),\quad
\phi(\theta)\to \phi(e^{-i\alpha\gamma_5}\theta),\quad
\tilde\phi(\theta)\to \tilde\phi(e^{-i\alpha\gamma_5}\theta);\nonumber\\
&&U(1)_2:\quad
W(\theta) \to W(\theta),\quad
\phi(\theta)\to e^{i\beta}\phi(\theta),\quad
\tilde\phi(\theta)\to e^{i\beta}\tilde\phi(\theta).
\end{eqnarray}

\noindent
that in components are written as

\begin{eqnarray}\label{symmetry}
U(1)_1: && A_\mu \to A_\mu;\qquad \varphi \to \varphi;\qquad
\tilde\varphi\to\tilde\varphi;\nonumber\\
&&\lambda\to e^{i\alpha\gamma_5}\lambda;\qquad
\Psi\to e^{-i\alpha\gamma_5}\Psi.\nonumber\\
\nonumber\\
U(1)_2: && A_\mu \to A_\mu;\qquad \varphi \to e^{i\beta}\varphi;\qquad
\tilde\varphi\to e^{i\beta}\tilde\varphi;\nonumber\\
&&\lambda\to\lambda;\qquad
\Psi\to e^{i\beta\gamma_5}\Psi.
\end{eqnarray}

The conservation of corresponding currents

\begin{eqnarray}
&& J^\mu_1 = \bar\lambda^a (1 +\gamma_5) \gamma^\mu \lambda^a
+\sum\limits_A \bar\Psi_A \gamma^\mu \gamma_5 \Psi_A; \nonumber\\
&& J^\mu_2 = -\sum\limits_A \bar\Psi_A \gamma^\mu \gamma_5 \Psi_A
-i \sum\limits_A \Big(\varphi_A^{*}
D^\mu \varphi_A - D^\mu\varphi_A^{*} \varphi_A
+ \tilde\varphi_A^{*}
D^\mu \tilde \varphi_A - D^\mu\tilde \varphi_A^{*} \tilde\varphi_A \Big).
\end{eqnarray}

\noindent
is destroyed at the quantum level by anomalies. In the perturbation theory

\begin{eqnarray}\label{curr}
&&\partial_\mu J^\mu_1 = (-N_f + N_c) \frac{1}{16\pi^2}
\varepsilon^{\mu\nu\alpha\beta} \mbox{tr} F_{\mu\nu} F_{\alpha\beta}
=(N_f-N_c)\frac{1}{16\pi^2} \mbox{Im\ tr}\int d^2\theta\ W^2;
\nonumber\\
&&\partial_\mu J^\mu_2 = N_f \frac{1}{16\pi^2}
\varepsilon^{\mu\nu\alpha\beta} \mbox{tr} F_{\mu\nu} F_{\alpha\beta}
=- N_f\frac{1}{16\pi^2} \mbox{Im\ tr}\int d^2\theta\ W^2.
\end{eqnarray}

Nevertheless, it is possible to construct an anomaly free symmetry. Really,
>from (\ref{curr}) we conclude, that

\begin{eqnarray}
&&J^\mu_R \equiv J_1^\mu + \frac{N_f-N_c}{N_f} J_2^\mu
= \bar\lambda^a (1+\gamma_5) \gamma^\mu \lambda^a
+ \frac{N_c}{N_f} \sum\limits_A \bar\Psi_A \gamma^\mu \gamma_5 \Psi_A
\nonumber\\
&&-i \sum\limits_A \left(1-\frac{N_c}{N_f}\right) \Big(\varphi_A^{*}
D^\mu \varphi_A - D^\mu\varphi_A^{*} \varphi_A
+ \tilde\varphi_A^{*}
D^\mu \tilde \varphi_A - D^\mu\tilde \varphi_A^{*} \tilde\varphi_A \Big)
\end{eqnarray}

\noindent
is conserved even at the quantum level.

This current is produced by the transformations

\begin{eqnarray}
U(1)_R: &&
W(\theta) \to e^{i\alpha_R} W \Big(e^{-i\alpha_R\gamma_5}\theta\Big);
\nonumber\\
&& \phi(\theta) \to \mbox{exp}\left(i\alpha_R\frac{N_f-N_c}{N_f}\right)\phi
\Big(e^{-i\alpha_R\gamma_5}\theta\Big);
\nonumber\\
&& \tilde \phi(\theta) \to \mbox{exp}
\left(i\alpha_R\frac{N_f-N_c}{N_f} \right)
\tilde\phi \Big(e^{-i\alpha_R\gamma_5}\theta\Big).
\end{eqnarray}

Below we will also use the combination of $U(1)_1$ and $U(1)_2$ with
$\beta=x\alpha$ in (\ref{symmetry}), i.e.

\begin{eqnarray}
U(1)_x: &&W(\theta) \to e^{i\alpha} W(e^{-i\alpha\gamma_5}\theta);\nonumber\\
&&\phi(\theta)\to e^{ix\alpha} \phi(e^{-i\alpha\gamma_5}\theta);\nonumber\\
&&\tilde\phi(\theta)\to e^{ix\alpha} \tilde\phi(e^{-i\alpha\gamma_5}\theta).
\end{eqnarray}

\noindent
where $x$ is an arbitrary constant.

In particular, for $x=(N_f-N_c)/N_f$ we obtain $U(1)_R$ transformations;
for $x=0$ - $U(1)_1$ and for $x \to \infty$ (after redefinition
$\alpha \to \alpha/x$) $U(1)_2$.

The corresponding current is

\begin{eqnarray}
&&J^\mu_x \equiv J_1^\mu + x J_2^\mu
= \bar\lambda^a (1+\gamma_5) \gamma^\mu \lambda^a
+ (1-x) \sum\limits_A \bar\Psi_A \gamma^\mu \gamma_5 \Psi_A
\nonumber\\
&&-i \sum\limits_A x \Big(\varphi_A^{*}
D^\mu \varphi_A - D^\mu\varphi_A^{*} \varphi_A
+ \tilde\varphi_A^{*}
D^\mu \tilde \varphi_A - D^\mu\tilde \varphi_A^{*} \tilde\varphi_A \Big).
\end{eqnarray}

\noindent
In the perturbation theory

\begin{eqnarray}\label{pertanomaly}
&&\partial_\mu J^\mu_x = \Big(-N_f + N_c + x N_f\Big) \frac{1}{16\pi^2}
\varepsilon^{\mu\nu\alpha\beta} \mbox{tr} F_{\mu\nu} F_{\alpha\beta}
\nonumber\\
&&\qquad\qquad\qquad\qquad\qquad\qquad
= \Big(N_f - N_c - x N_f\Big) \frac{1}{16\pi^2}
\mbox{Im\ tr}\int d^2\theta\ W^2 .
\end{eqnarray}


\subsection{N=2 sypersymmetry}\label{N2}
\hspace{\parindent}

In the superspace N=2 sypersymmetric Yang-Mills theory is described by
the action

\begin{equation}\label{action1}
S = \frac{1}{32\pi} \mbox{tr}\ \mbox{Im} \Big(
\tau \int d^4 d^2\theta_1 d^2\theta_2
\frac{1}{2} \Phi^2\Big)
\end{equation}

\noindent
where

\begin{eqnarray}
&&\Phi(y,\theta_1,\theta_2) = \phi(y,\theta_1) - i \bar\theta_2(1+\gamma_5)
W(y,\theta_1) + \frac{1}{2} \bar\theta_2 (1+\gamma_5) \theta_2 G(y,\theta_1);
\nonumber\\
&&y^\mu = x^\mu
+{\displaystyle \frac{i}{2}} \bar\theta_i \gamma^\mu\gamma_5\theta_i;
\nonumber\\
&&G(y,\theta_1) = \frac{1}{2} \int d^2\bar \theta_1 e^{2V} \phi^{+} e^{-2V}.
\end{eqnarray}

\noindent
where $\phi$ and $W$ were defined in (\ref{notat}).

Denoting $\varphi \equiv P+iS$, $\psi_1\equiv \lambda$ and
$\psi_2\equiv\psi$, we find that in components

\begin{eqnarray}
&& S = \frac{1}{e^2}\mbox{tr} \int d^4x \Big(
- \frac{1}{4} F_{\mu\nu}^2 - i\bar \psi_i \gamma^\mu D_\mu \psi_i
+\frac{1}{2} (D_\mu P)^2 + \frac{1}{2} (D_\mu S)^2 -\frac{1}{2} [P,S]^2
\nonumber\\
&&
- i\epsilon_{ij} \{\bar \psi_i, \gamma_5\psi_j\} P
- \epsilon_{ij} \{\bar \psi_i, \psi_j\} S
+\frac{\theta e^2}{32\pi^2} F_{\mu\nu}\tilde F_{\mu\nu}
\Big)
\end{eqnarray}

In this paper we will consider only the case of SU(n) gauge group.

The action (\ref{action1}) is invariant under the transformations

\begin{equation}
U(1)_R: \quad\Phi(\theta) \to e^{2i\alpha}\Phi(e^{-i\alpha\gamma_5}\theta)
\end{equation}

\noindent
In components they are written as

\begin{equation}
\varphi \to e^{2i\alpha} \varphi;\qquad
\psi_i \to e^{i\alpha\gamma_5} \psi_i;\qquad
A_\mu \to A_\mu
\end{equation}

So, R-symmetry leads to the chiral transformations for fermions. Using
the expression for the axial anomaly we find that in the perturbation
theory for SU(n) gauge group

\begin{equation}\label{anomal}
\langle\partial_\mu j_R^\mu\rangle_{pert}
= \frac{n}{16\pi^2} \varepsilon^{\mu\nu\alpha\beta}
\mbox{tr} F_{\mu\nu} F_{\alpha\beta}
=- \frac{n}{32\pi^2}
\mbox{Im}\ \mbox{tr}\int d^2\theta_1 d^2\theta_2 \Phi^2,
\end{equation}

\noindent
where

\begin{equation}
j_R^\mu = \bar\psi_i^a \gamma^\mu \gamma_5 \psi_i^a - 4 D^\mu P^a S^a
+ 4 D^\mu S^a P^a.
\end{equation}

Below we will see, that (\ref{anomal}) is no longer valid beyond the frames
of the perturbation theory. Although the existence of instanton corrections
was predicted rather long ago \cite{shifmanomaly}, it is much better to
have an exact result. Its derivation requires information about the
vacuum structure and low energy limit of the theory. Here we would like to
remind some key points.

The classical potential for the scalar superfield component $\varphi$
is given by

\begin{equation}
- \frac{1}{4}[\varphi,\varphi^{+}]_a^2 = \frac{1}{2}[P,S]_a^2.
\end{equation}

\noindent
It leads to a continuous family of unequivalent ground states, which
constitutes the classical moduli space $M_0$. In order to characterize
$M_0$ we note, that one can always rotate $\varphi$ into Cartan sub-algebra

\begin{equation}
\varphi=\sum\limits_{k=1}^{r} a_k H_h.
\end{equation}

\noindent
Here $r$ denotes the rank of gauge group G. Below we will consider only
G=SU(n), so that r=n-1. In the generic point of $M_0$ it is spontaneously
broken down to $U(1)^{n-1}$.

The Cartan sub-algebra variables $a_i$ are not gauge invariant, and,
therefore, one should introduce other variables for parametrizing the
classical moduli space. It can be done in the following way:

Let us consider

\begin{equation}
W_{A_{n-1}}\equiv \langle\mbox{det} (x{\bf 1}-\varphi)\rangle
\end{equation}

\noindent
whose coefficients are gauge invariant. If (in the case of SU(n))

\begin{equation}
\varphi=\mbox{diag}(a_1,a_2,\ldots,a_n), \qquad \sum\limits_i a_i = 0,
\end{equation}

\noindent
we find that classically

\begin{equation}
W_{A_{n-1}} = x^n + x^{n-2}\sum\limits_{i<j}a_i a_j
- x^{n-3}\sum\limits_{i<j<k}a_i a_j a_k+\ldots+
(-1)^n \prod \limits_i a_i.
\end{equation}

\noindent
>From the other hand

\begin{eqnarray}
&&W_{A_{n-1}}=\langle x^n det\Big({\bf 1}-\frac{\varphi}{x}\Big) \rangle =
x^n \langle \mbox{exp}\left[\mbox{tr}\ \ln\Big({\bf 1}-\frac{\varphi}{x}\Big)
\right]\rangle=\nonumber\\
&&= x^n + x^{n-2} \frac{1}{2} \langle\mbox{tr} \varphi^2\rangle
- x^{n-3} \frac{1}{3} \langle \mbox{tr} \varphi^3\rangle + O(x^{n-4})
\equiv
x^n + \sum\limits_{k=1}^{n-1} x^{n-1-k} u_k(a).
\end{eqnarray}

Therefore, the gauge invariant description of the theory can be made
in terms of

\begin{equation}
u_k(a) = \prod\limits_{n_1<n_2<\ldots<n_k} a_{n_1} a_{n_2}\ldots a_{n_k}.
\end{equation}

\noindent
In particular,

\begin{equation}
u_1 = \frac{1}{2}\langle\mbox{tr} \varphi^2\rangle;\qquad
u_2 = - \frac{1}{3}\langle\mbox{tr} \varphi^3\rangle.
\end{equation}

\noindent
In the cases of SU(2) and SU(3) it is easy to see, that

\begin{eqnarray}
SU(2):&& W_{A_1}=x^2-u\vphantom{\frac{1}{2}};\nonumber\\
&& \quad u_1 \equiv u = \frac{1}{2}\mbox{tr}\varphi^2 =\frac{1}{2} a^2;
\nonumber\\
SU(3):&& W_{A_2}=x^3-xu-v\vphantom{\frac{1}{2}};\nonumber\\
&& \quad u_1\equiv u=\frac{1}{2}\mbox{tr}\langle\varphi^2\rangle
=-\sum\limits_{i<j}a_i a_j = a_1^2+a_2^2+a_1 a_2;
\nonumber\\
&& \quad u_2\equiv v=-\frac{1}{3}\mbox{tr}
\langle\varphi^3\rangle = - a_1 a_2 a_3 = a_1 a_2 (a_1+a_2).
\end{eqnarray}

As we mentioned above, in the low energy limit the theory is described
by $r=n-1$ N=2 abelian superfields $\Phi_i$. N=2 supersymmetry constrains
the form of effective action to be

\begin{equation}\label{N2action}
\Gamma =
\frac{1}{32\pi} \mbox{Im} \int d^4x d^2\theta_1 d^2\theta_2 F(\Phi_i)
\end{equation}

\noindent
where $F$, called prepotential, depends only on $\Phi$ and not on $\Phi^{+}$.

The low energy effective action was shown \cite{seiberg} to be invariant
under duality transformations

\begin{eqnarray}\label{dual}
&&F(\Phi) \to F_D(\Phi_D) = F(\Phi)
- \Phi_i \Phi_D^i\Big|_{\Phi=\Phi(\Phi_D)};
\nonumber\\
&& \Phi \to \Phi_D^i = \frac{\partial F(\Phi)}{\partial \Phi_i}.
\end{eqnarray}

Vacuum expectation values of dual superfields we denote as $a_D^i$.

The explicit form of $a_i$ and $a_D^i$ is usually found by Seiberg-Witten
elliptic curve method \cite{seiberg,klemm1}. However, there is another
approach. Let us note, that

\begin{equation}
\vec a = \left(\begin{array}{c}a_i\\ a_D^i\end{array}\right)
\end{equation}

\noindent
satisfies a system of second order differential equations (so called
Picard-Fuchs equations). Its explicit form was found to be

\begin{equation}\label{su2}
\Big(4(u^2-\Lambda^4)\partial_u^2 + 1\Big) \vec a = 0
\end{equation}

\noindent
for the case of SU(2) \cite{bilal} and

\begin{eqnarray}\label{su3}
&&\Big((27 \Lambda^6-4u^3-27 v^2)\partial_u^2 -12u^2v\partial_u\partial_v
-3uv\partial_v-u\Big)\vec a =0\nonumber\\
&&\Big((27 \Lambda^6-4u^3-27 v^2)\partial_v^2 -36 uv\partial_u\partial_v
-9v\partial_v-3\Big)\vec a =0
\end{eqnarray}

\noindent
for the SU(3) \cite{klemm2} gauge group. (Here $\Lambda$ is the instanton
generated scale.)

Below we will see, that Picard-Fuchs equations play a crucial role in
the consideration of anomalies. Actually, they assure, that the relation
between perturbative and exact anomalies is satisfied.

In the SU(2) case one can easily solve (\ref{su2}) and finds, that
taking into account perturbative assimptotics

\begin{eqnarray}\label{sw}
&&a(u)=\frac{\sqrt{2}}{\pi}\int\limits_{-\Lambda^2}^{\Lambda^2} dx
\frac{\sqrt{x-u}}{\sqrt{x^2-\Lambda^4}};
\nonumber\\
&&a_D(u)=\frac{\sqrt{2}}{\pi}\int\limits_{\Lambda^2}^{u} dx
\frac{\sqrt{x-u}}{\sqrt{x^2-\Lambda^4}}.
\end{eqnarray}

The prepotential $F$ can be then found by

\begin{equation}
\frac{dF}{du} = a_D \frac{da}{du}
\end{equation}

\noindent
and its perturbative asymptotics.


\section{Instanton contributions to anomalies and the structure of
the effective potential}\label{instanton}

\subsection{N=2 SUSY SU(2) Yang-Mills theory}
\hspace{\parindent}

First we would like to discuss how instantons contribute to anomalies.
On the one hand anomalies can be defined from the effective action and,
therefore, instanton corrections to the effective action lead to the
instanton corrections to anomalies. However, in this section we will
try to investigate the mechanism of their appearance and show, that
nonperturbative contributions arise due to the noninvariance of collective
coordinate measure. The developed approach extends the method, presented in
\cite{shifmanomaly}. However, comparing the results with the form of the
effective action allows to predict the structure of nonperturbative
superpotential, which will be used below.\footnote{
Of course, this structure can be obtain from dimensional arguments, but the
presented approach seems more relevant in the paper, devoted to anomalies.}

Nonperturbative contributions to the effective action are obtained by
the expansion of the generating functional near the instanton solution
\cite{thooft}

\begin{equation}\label{genfun}
\Delta \Gamma = \frac{1}{Z_0} \int d\mu\ D\phi\
\mbox{exp}\Big(-S[\phi_{0}+\phi]\Big).
\end{equation}

\noindent
Here $\phi$ denotes the whole set of fields, $\phi_0$ is a classical
instanton solution and $d\mu$ is a collective coordinate measure.

At the one-instanton level there are 8 bosonic zero modes, due to the
invariances under 4 translations (the corresponding collective coordinates
are $a^\mu$), rescaling ($\rho$) and 3 gauge transformations ($\omega$).
Moreover, N=2 supersymmetry adds 4 supersymmetries ($\epsilon_i$) and
4 superconformal transformations ($\beta_i$) \cite{cordes}. If the scalar
field has nonzero vacuum expectation value, the superconformal modes
are lifted due to the conformal symmetry breaking. Nevertheless, we still
keep the integration over the corresponding parameters in the instanton
measure following \cite{affleck,shifman,yung}.

The final expression for the one-instanton measure \cite{yung} is

\begin{equation}
d\mu = \mbox{const}
\int d^4a \frac{d\rho}{\rho^5} \frac{d^3\omega}{2\pi^2} M^8 \rho^8
\times \frac{1}{M^4 \rho^4}
\int d^2\epsilon_1 d^2\epsilon_2 d^2\bar\beta_1 d^2\bar\beta_2
\end{equation}

A next step to obtain nonperturbative corrections is the calculation
of the exponent in the constant field limit \cite{callan,shifman}.
However, it can be omitted, because we are going to investigate only the
general structure of instanton corrections.

Really, let us perform $U(1)_R$-transformations in the generating functional
(\ref{genfun}). Because

\begin{eqnarray}
&&\epsilon_i \to e^{i\alpha\gamma_5}\epsilon_i;
\qquad\qquad\ \rho \to \rho;\nonumber\\
&&\beta_i \to e^{i\alpha\gamma_5}\beta_i;
\qquad\qquad a^\mu \to a^\mu;\nonumber\\
&&\qquad\qquad\quad\delta\Phi \to e^{2i\alpha}\Phi
\end{eqnarray}

\noindent
the collective coordinate measure is not invariant. It is easy to see,
that the exponent in (\ref{genfun}), being a function of collective
coordinates,  is not invariant too. For example, the scalar field
contribution $4\pi^2 \rho^2 \Phi^2$ \cite{thooft} has evidently nontrivial
transformation law.

However, the invariance of the exponent can be easily restored by
additional variable substitution
\footnote{Note, that the transformations of $\theta$ and $x^\mu$ are not
independent}

\begin{equation}
\theta \to e^{-i\alpha\gamma_5}\theta;\qquad
\rho \to e^{-2i\alpha} \rho;\qquad
a^\mu \to e^{-2i\alpha} a^\mu,
\end{equation}

\noindent
because the overall transformations

\begin{eqnarray}\label{N2transformations}
&&x^\mu \to e^{-2i\alpha} x^\mu;\qquad\qquad
\theta \to e^{-2 i\alpha\gamma_5}\theta;\nonumber\\
&&a^\mu \to e^{-2i\alpha} a^\mu;\qquad\qquad
(1+\gamma_5)\epsilon \to e^{i\alpha}(1+\gamma_5)\epsilon;\nonumber\\
&&\rho \to e^{-2i\alpha} \rho;\qquad\qquad\quad
(1-\gamma_5)\beta \to e^{-i\alpha}(1-\gamma_5)\beta
\end{eqnarray}

\noindent
do not effect any dimensionless function of collective coordinates
\footnote{Of course, our method is in a deep connection with
dimensional arguments, although it does not completely repeat them.
For example, $\theta$-transformation law does not correspond to its
dimension.}.

Thus, only the instanton measure is transformed nontrivially under
(\ref{N2transformations})

\begin{equation}
d\mu \to \Big(e^{-2i\alpha}\Big)^4 \Big(e^{-i\alpha}\Big)^4
\Big(e^{i\alpha}\Big)^4 d\mu
= e^{-8i\alpha} d\mu.
\end{equation}

For n-instanton contribution we should consider the limit, where a
multiinstanton solution is presented as a sum of n instantons, distant
>from each other. Then we immediately conclude, that

\begin{equation}
d\mu^{(n)} \to e^{-8in\alpha} d\mu^{(n)}.
\end{equation}

Taking into account that

\begin{equation}
\int d^4x d^2\theta_1 d^2\theta_2
\end{equation}

\noindent
(which is present in the prepotential definition) remains invariant
under (\ref{N2transformations}) we find

\begin{equation}\label{a1}
\langle\partial_\mu j_R^\mu\rangle =
-\frac{\partial \Gamma}{\partial \alpha}\Big|_{\alpha=0}=
\sum\limits_{n=0}^\infty \frac{1}{32\pi} \mbox{Im} \Big(8in
\int d^2\theta_1 d^2\theta_2 \Delta F^{(n)}\Big),
\end{equation}

\noindent
where $\Delta F^{(n)}$ is n-instanton contribution to the prepotential.

On the other hand, transforming (\ref{N2action}) we obtain

\begin{equation}
\langle\partial_\mu j_R^\mu\rangle =
- \frac{1}{32\pi} \mbox{Im} \int d^2\theta_1 d^2\theta_2
\Big(2i\Phi\frac{\partial}{\partial \Phi}- 4i\Big)
\sum\limits_{n=0}^\infty\Delta F^{(n)}.
\end{equation}

\noindent
Solving the equation

\begin{equation}
\Phi \frac{\partial}{\partial \Phi} \Delta F^{(n)}
=\Big(-4n+2\Big) \Delta F^{(n)}
\end{equation}

\noindent
we obtain the final structure of the nonperturbative anomaly to be

\begin{equation}
\langle\partial_\mu j_R^\mu\rangle =
- \frac{1}{16\pi^2} \mbox{Im}\int d^4x d^2\theta_1 d^2\theta_2
\sum\limits_{n=0}^\infty c_n \Phi^{2(-2n+1)},
\end{equation}

\noindent
where $c_n\ \epsilon\ \mbox{Re}$ and $c_0=1$


\subsection{N=1 SUSY Yang-Mills with matter}
\hspace{\parindent}

Now let us consider N=1 supersymmetric $SU(N_c)$ Yang-Mills theory
with $N_f$ matter supermultiplets and find the general {\it possible}
structure of instanton corrections to superpotential and anomalies.

The effective Lagrangian can be split into the following parts
\cite{veneziano}

\begin{equation}
L_{eff} = L_{k} +  L_{a} + L_m,
\end{equation}

\noindent
where

\begin{eqnarray}
&& L_k = \int d^4\theta\ K(S, S^{*},\Phi, \Phi^{*}); \nonumber\\
&& L_a = \mbox{Re} \int d^2\theta\ w(S,\Phi)
\end{eqnarray}

\noindent
and $S\equiv\mbox{tr} W^2$

Here $L_k$ denotes kinetic terms, that do not contribute to the anomaly,
$L_{a}$ is a holomorphic part of the superpotential and $L_m$ is a mass
term. Below we will consider only massless case ($L_m=0$). Therefore,
the only nontrivial contribution to anomalies comes from $L_a$ and it is
the only part, that we are able to investigate. (Our method can not give
any information about a possible kinetic term.)

As above we will calculate anomalies by 2 different ways and compare
the results. The action is invariant under $U(1)_1\times U(1)_2$
group. However, it is more convenient to investigate the anomaly of
$U(1)_x$ symmetry, constructed in Section \ref{n1}.

Performing $U(1)_x$ transformation in the effective action we obtain

\begin{equation}\label{N1anomaly1}
\langle\partial_\mu J^\mu_x\rangle =-\frac{\partial\Gamma}{\partial\alpha}
=- \mbox{Im} \int d^2\theta \Big(2w-2\frac{\partial w}{\partial S} S
- x \frac{\partial w}{\partial v} v\Big),
\end{equation}

\noindent
where we substituted $\phi$ and $\tilde\phi$ by their vacuum expectation
values $v$. (For simplicity we assume, that all $v_i$ are equal; a brief
review of the moduli space structure is given in the Appendix \ref{moduli}.)

On the other hand, the anomaly can be found from the transformation law of
the collective coordinate measure.

At the one-instanton level in this case there are 8 bose zero modes (exactly
as above), $2 N_c$ gluino zero modes (corresponding to supersymmetric
($\epsilon_a$) and superconformal ($\beta_a$) transformations) and $2N_f$
zero modes for matter multiplets (supersymmetry $\epsilon_A$). Each zero
mode should be removed by integration over the corresponding collective
coordinate. The measure is written as \cite{cordes}

\begin{eqnarray}\label{measure}
&&d\mu=\nonumber\\
&&=\mbox{const}
\int d^4a \frac{d\rho}{\rho^5} (M\rho)^{4 N_c} d(\mbox{gauge})
\frac{1}{M^{N_c+N_f}\rho^{N_c}}
\prod\limits_{a=1}^{N_c} d\epsilon_a d\bar\beta_a
\prod\limits_{A=1}^{N_f} \frac{d\epsilon_A d\tilde\epsilon_A}{\rho^2 v^2}
\mbox{exp}\left(-\frac{8\pi^2}{e^2} \right) \nonumber\\
&&=\mbox{const} \Lambda^{3N_c-N_f}
\int d^4a d\rho \rho^{3 N_c- 2 N_f-5} \frac{1}{v^{2 N_f}} d(\mbox{gauge})
\prod\limits_{a=1}^{N_c} d\epsilon_a d\bar\beta_a
\prod\limits_{A=1}^{N_f} d\epsilon_A d\tilde\epsilon_A,
\end{eqnarray}

\noindent
where we take into account normalization of all zero modes. The gauge part
and constant factors are written only schematically, because they are not
important in our discussion. As above we need not know the explicit form of
the action in the constant field limit. We should only emphasize, that it
is a dimensionless function of collective coordinates, $\phi$ and, in
principle, $W$. Of course, it is not invariant under $U(1)_x$-transformations

\begin{eqnarray}
&&W  \to e^{i\alpha\gamma_5}W;\qquad\quad\ \
\theta \to e^{-i\alpha\gamma_5}\theta
\nonumber\\
&&\phi \to e^{ix\alpha}\phi;\qquad\qquad\quad
\tilde\phi \to e^{ix\alpha}\tilde\phi;\nonumber\\
&&\epsilon_a \to e^{i\alpha\gamma_5}\epsilon_a;\qquad\qquad\
\beta_a \to e^{i\alpha\gamma_5}\beta_a;\nonumber\\
&&\epsilon_A \to e^{i(x-1)\alpha\gamma_5}\epsilon_A;\qquad
\tilde\epsilon_A \to e^{i(x-1)\alpha\gamma_5}\tilde\epsilon_A;
\nonumber\\
&&\rho \to \rho;\qquad\qquad\qquad\ \ \ a^\mu{} \to a^\mu
\end{eqnarray}

\noindent
as above.

Similarly to N=2 supersymmetric Yang-Mills theory we perform an additional
substitution

\begin{eqnarray}\label{compens}
&&\theta \to e^{-i\alpha\gamma_5}\theta;\qquad\qquad\quad\ \
x^\mu \to e^{-2i\alpha} x^\mu;\qquad\nonumber\\
&&\epsilon_A \to e^{-ix\alpha\gamma_5}\epsilon_A;\qquad\qquad\
\tilde\epsilon_A \to e^{-ix\alpha\gamma_5}\tilde\epsilon_A;
\qquad\nonumber\\
&&\rho \to e^{-2i\alpha} \rho;\qquad\qquad\qquad
a^\mu \to e^{-2i\alpha} a^\mu,\qquad\nonumber\\
\end{eqnarray}

\noindent
so that the final transformations

\begin{eqnarray}\label{overall}
&&(1+\gamma_5)\epsilon_a \to e^{i\alpha}(1+\gamma_5)\epsilon_a;
\qquad\qquad\ \
(1-\gamma_5)\beta_a \to e^{-i\alpha}(1-\gamma_5)\beta_a;\qquad\nonumber\\
&&(1+\gamma_5)\epsilon_A \to e^{-i\alpha}(1+\gamma_5)\epsilon_A;\qquad\qquad
(1+\gamma_5)\tilde\epsilon_A \to e^{-i\alpha}(1+\gamma_5)
\tilde\epsilon_A;\qquad\nonumber\\
&&\rho \to e^{-2i\alpha} \rho;\qquad\qquad\qquad
a^\mu \to e^{-2i\alpha} a^\mu;\qquad\qquad\qquad
\theta \to e^{-2i\alpha\gamma_5}\theta\qquad
\end{eqnarray}

\noindent
(except for $\theta$) correspond to dimension of the fields. The
dimensionless action would have been invariant, if we had made additional
rotation

\begin{equation}\label{lack}
v \to e^{i(2-x)\alpha} v;\qquad
W \to e^{2i\alpha\gamma_5} W.
\end{equation}

\noindent
However, we can not make it because $v$ and $W$ are not collective
coordinates (and, therefore, integration variables). It means, that
under (\ref{overall})

\begin{eqnarray}
&&S(v,W) \to S(e^{i(x-2)\alpha} v, e^{-2i\alpha\gamma_5} W);\nonumber\\
&&d\mu(v) \to \mbox{exp}\left[i\alpha\Big(-2(3N_c-2N_f) -2N_fx + 2N_f(x-2)
- 2N_f\Big)\right] d\mu(e^{i(x-2)\alpha}v)  \nonumber\\
&&=\mbox{exp}\left[i\alpha\Big(-2(3N_c-N_f)\Big)\right]
d\mu(e^{i(x-2)\alpha}v).
\end{eqnarray}

\noindent
It is quite evident, that the $n$-instanton collective coordinate
measure is transformed as

\begin{equation}\label{measurecontribution}
d\mu(v) \to \mbox{exp}\left[i\alpha\Big(-2n(3N_c-N_f) \Big)
\right] d\mu(e^{i(x-2)\alpha}v).
\end{equation}

\noindent
Moreover, we should also perform the inverse substitution in the remaining
integral (see the definition of the superpotential)

\begin{equation}\label{lastint}
\int d^4x d^2\theta \to e^{4i\alpha}\int d^4x d^2\theta,
\end{equation}

\noindent
so that finally from (\ref{lack}), (\ref{measurecontribution}) and
(\ref{lastint}) we conclude, that

\begin{equation}
w(v,W) \to \mbox{exp}\left[i\alpha\Big(-2n(3N_c-N_f)+4 \Big)\right]
w(e^{i(x-2)\alpha} v, e^{-2i\alpha\gamma_5} W).
\end{equation}

\noindent
Taking into account that the action contains only $(1+\gamma_5) W$, we find
the anomaly to be

\begin{eqnarray}\label{N1anomaly2}
&&\langle\partial_\mu J^\mu_R\rangle
=-\left. \frac{\partial\Gamma}{\partial\alpha}\right|_{\alpha=0}=
\nonumber\\
&&\qquad\qquad =\mbox{Im} \int d^2\theta \left(- 2n(3N_c-N_f)
+ (-2+x) v\frac{\partial}{\partial v} - 2 W\frac{\partial}{\partial W}
+ 4 \right) w. \qquad
\end{eqnarray}

\noindent
Comparing (\ref{N1anomaly1}) and (\ref{N1anomaly2}), we obtain the
following equation for $n$-instanton contribution to the superpotential:

\begin{equation}
\Big(2v\frac{\partial}{\partial v}
+ 3 W \frac{\partial}{\partial W} - 6 \Big) w^{(n)}
= - 2 n (3N_c-N_f) w^{(n)}.
\end{equation}

\noindent
It is easily verified, that the solution is

\begin{equation}\label{superpotential0}
w^{(n)}
=  W^2 g_n\left(\frac{v^3}{W^2}\right)
\left(\frac{\Lambda}{v}\right)^{n(3N_c-N_f)}
=  S g_n\left(\frac{v^3}{S}\right)
\left(\frac{\Lambda}{v}\right)^{n(3N_c-N_f)}
\end{equation}

\noindent
where $g_n$ is an arbitrary function. Its explicit form will be found below
>from the relation between perturbative and exact anomalies.

Of course, the result (\ref{superpotential0}) is in a complete agreement
with dimensional arguments and does not depend on the particular choice of
symmetry (i.e. $x$).


\section{The relation between perturbative and exact anomalies}
\label{rel}

\subsection{Exact anomaly in the N=2 SUSY SU(2) case}
\label{repeat}
\hspace{\parindent}

Let us briefly remind the calculation of exact R-anomaly following
\cite{matone,bellisai}. The anomaly can be found from the effective action

\begin{equation}
\langle\partial_\mu j_R^\mu\rangle =
-\frac{\partial \Gamma}{\partial \alpha}\Big|_{\alpha=0} =
-\frac{1}{32\pi} \mbox{Im} \frac{\partial}{\partial\alpha}
\int d^4x d^2\theta_1 d^2\theta_2 F\Big(e^{2i\alpha}
\Phi(e^{-4i\alpha\gamma_5} \theta)\Big)\Big|_{\alpha=0}.
\end{equation}

Taking into account, that ${\displaystyle \int d\theta}$ is actually a
differentiation over the anticommuting variables and it is possible to
perform the additional rotation $\theta \to e^{i\alpha\gamma_5}\theta$, 
we obtain

\begin{eqnarray}
&&\langle\partial_\mu j_R^\mu\rangle =
-\frac{1}{32\pi} \mbox{Im} \frac{\partial}{\partial\alpha}
\int d^4x d^2\theta_1 d^2\theta_2 e^{-4i\alpha}
F\Big(e^{2i\alpha}\Phi(\theta)\Big)\Big|_{\alpha=0} =\nonumber\\
&&=\frac{1}{16\pi} \mbox{Re} \int d^2\theta_1 d^2\theta_2
\Big(2 F(\Phi) - \frac{\partial F}{\partial \Phi}\Phi \Big)
= \frac{1}{16\pi} \mbox{Re} \int d^2\theta_1 d^2\theta_2
\Big(F+F_D \Big).\qquad
\end{eqnarray}

This expression can be found explicitly, really

\begin{eqnarray}
&&\frac{d}{du}\Big(F+F_D \Big)
=\frac{dF}{da} \frac{da}{du} + \frac{dF_D}{da_D} \frac{da_D}{du}
=a_D \frac{da}{du} - a \frac{da_D}{du};\nonumber\\
&&\frac{d^2}{du^2}\Big(F+F_D \Big) =
a_D \frac{d^2 a}{du^2} - a \frac{d^2 a_D}{du^2}=0,
\end{eqnarray}

\noindent
where we used (\ref{su2}).

Therefore, taking into account perturbative asymptotics finally we have

\begin{equation}
F+F_D= \mbox{const}\ u =\frac{2i}{\pi} u,
\end{equation}

\noindent
so that the anomaly can be written as

\begin{equation}\label{npa}
\langle\partial_\mu j_R^\mu\rangle =
- \frac{1}{8\pi^2} \mbox{Im} \int d^2\theta_1 d^2\theta_2 u,
\end{equation}

\noindent
while in the perturbation theory

\begin{equation}\label{pa}
\mbox{\bf A}\equiv \langle\partial_\mu j_R^\mu\rangle_{pert} =
- \frac{1}{16\pi^2} \mbox{Im}\ \mbox{tr}\int d^2\theta_1 d^2\theta_2 \Phi^2.
\end{equation}

\subsection{Derivation}
\label{condition}
\hspace{\parindent}

Of course, the expressions (\ref{npa}) and (\ref{pa}) are quite different.
The former is a series over $\Lambda^4$ produced by instanton contributions.
In particular, taking into account one instanton correction we have
\cite{finnell,yung}

\begin{equation}\label{correct}
\langle\partial_\mu j_R^\mu\rangle
= - \frac{1}{16\pi^2} \mbox{Im}\ \mbox{tr}\int d^2\theta_1 d^2\theta_2
\left[\Phi^2 +\frac{\Lambda^4}{2\Phi^2} +O(\Lambda^8)\right],
\end{equation}

\noindent
that in components can be written as

\begin{eqnarray}
&&\langle\partial_\mu j_R^\mu\rangle
=\frac{e^2}{4\pi^2} \Big(1 + \frac{3\Lambda^4}{2 e^4 \varphi^4}\Big)
F_{\mu\nu} \tilde F^{\mu\nu}
-\frac{3\Lambda^4}{e^2 \pi^2 \varphi^5}F_{\mu\nu}
\bar\Psi_D \Sigma_{\mu\nu}\gamma_5 \Psi_D\nonumber\\
&&\qquad\qquad\qquad\qquad\qquad\qquad\qquad\qquad
+\frac{60\Lambda^4}{e^2 \pi^2 \varphi^6}
(\bar\Psi_D\Psi_D) (\bar\Psi_D\gamma_5 \Psi_D)+O(\Lambda^8),\qquad
\end{eqnarray}

\noindent
where

\begin{equation}
\tilde F^{\mu\nu} = \frac{1}{2} \varepsilon^{\mu\nu\alpha\beta}
F_{\alpha\beta}
\end{equation}

\noindent
and we introduced a Dirac spinor

\begin{equation}
\Psi_D = \frac{1}{2}(1+\gamma_5)\psi_1+\frac{1}{2}(1-\gamma_5)\psi_2.
\end{equation}

And nevertheless, nonperturbative result is only a vacuum expectation value
of the perturbative one, that in particular produces a natural solution of
anomalies cancellation problem in the realistic models.

This result is not unexpected. Really, performing, for example, chiral
transformation in the generating functional we have

\begin{eqnarray}
&&0 = \left.\frac{1}{Z} \frac{\delta Z}{\delta\alpha}\right|_{\alpha=0}
=\left.\frac{1}{Z} \frac{\delta}{\delta\alpha} \int DA D\bar\psi' D\psi'
\mbox{exp} \Big(i S - \partial_\mu\alpha j^\mu_5 \Big)\right|_{\alpha=0}
\nonumber\\
&&=\left.\frac{1}{Z} \frac{\delta}{\delta\alpha} \int DA D\bar\psi D\psi
\mbox{exp} \Big(i S - \partial_\mu\alpha j^\mu_5 - \alpha\mbox{\bf A}\Big)
\right|_{\alpha=0}
=\langle  \partial_\mu j^\mu_5 - \mbox{\bf A}  \rangle,
\end{eqnarray}

\noindent
where {\bf A} denotes the perturbative anomaly, produced by the measure
noninvariance \cite{fujikava}. Finally

\begin{equation}\label{relation}
\langle \partial_\mu j^\mu_5\rangle = \langle\mbox{\bf A}  \rangle.
\end{equation}

\noindent
(R-transformation are considered similarly).

It is just the relation, mentioned above. Of course, it is valid for a wide
range of models and is really a point to start with. Let us note, that the
derivation presented in Section \ref{repeat} essentially used the form of
exact results. So, we are tempted to reverse the arguments. In the next
section we will try to develop this approach.


\section{Consequences and results}
\label{section3}

\subsection{N=2 SUSY SU(n): Picard - Fuchs equations versus anomalies}
\label{n2susy}
\hspace{\parindent}

We will start the investigation of the SU(n) case with the relation
(\ref{relation}). It means that

\begin{equation}\label{anomaly1}
\langle\partial_\mu j_R^\mu\rangle =
- \frac{n}{16\pi^2} \mbox{Im} \int d^2\theta_1 d^2\theta_2 u_1.
\end{equation}

\noindent
>From the other hand

\begin{equation}\label{anomaly2}
\langle\partial_\mu j_R^\mu\rangle =
\frac{1}{16\pi} \mbox{Re} \int d^2\theta_1 d^2\theta_2
\Big(F+F_D \Big)
\end{equation}

\noindent
as above. Comparing (\ref{anomaly1}) and (\ref{anomaly2}) we find that

\begin{equation}
F+F_D= \frac{i n}{\pi} u_1
\end{equation}

\noindent
and, therefore,

\begin{equation}\label{deriv1}
\frac{\partial}{\partial u_k}\Big(F+F_D \Big)
=\frac{\partial F}{\partial a_i} \frac{\partial a_i}{\partial u_k}
+ \frac{\partial F_D}{\partial a_D^i} \frac{\partial a_D^i}{\partial u_k}
=a_D^i \frac{\partial a_i}{\partial u_k}
- a_i \frac{\partial a_D^i}{\partial u_k}=\frac{in}{\pi}\delta_{k1}.
\end{equation}

In the case of SU(2) gauge group

\begin{equation}
{\bf W} \equiv a_D \frac{da}{du} - a \frac{da_D}{du}
\end{equation}

\noindent
can be identified with the Wronsky determinant for a linear second order
differential equation, $a$ and $a_D$ being 2 its linear independent
solutions. The condition ${\bf W}=\mbox{const}$ constrains a form of
the equation to be

\begin{equation}
\Big(\partial_u^2 + L(u)\Big)
\left(\begin{array}{c}a\\a_D\end{array}\right)=0
\end{equation}

\noindent
due to the Liouville formula

\begin{equation}
{\bf W}(u) = {\bf W}(u_0) \mbox{exp}\Big(-\int\limits_{u_0}^u du' K(u')\Big),
\end{equation}

\noindent
where $K(u)$ is a coefficient of a first derivative term in the equation.

The SU(n) case can be considered similarly. Let us differentiate
(\ref{deriv1}) once more and contract the result with a symmetric
matrix $M_{km}(u)$

\begin{equation}\label{contr}
M_{km}\frac{\partial }{\partial u_m}
\Big(a_D^i \frac{\partial a_i}{\partial u_k}
- a_i \frac{\partial a_D^i}{\partial u_k}\Big)
=a_D^i M_{km}\frac{\partial^2 a_i}{\partial u_k \partial u_m}
- a_i M_{km}\frac{\partial^2 a_D^i}{\partial u_k \partial u_m}=0.
\end{equation}

\noindent
It means, that $a$ and $a_D$ should satisfy a system of the form

\begin{equation}\label{picard0}
\left(\delta_{ij} M_{km}(u)\frac{\partial^2}{\partial u_k \partial u_m}
+\delta_{ij} K_m(u)\frac{\partial}{\partial u_m} + L(u)_{ij}\right)
\left(\begin{array}{c} a_j\\a_D^j \end{array}\right) = 0.
\end{equation}

\noindent
Substituting it to (\ref{contr}) leads to

\begin{equation}
K_m\left(a_i \frac{\partial a_D^i}{\partial u_m}
- a_D^i \frac{\partial a_i}{\partial u_m} \right)=0,
\end{equation}

\noindent
so that using (\ref{deriv1}) we obtain the constrain $K_1(u)=0$.
Thus the Picard-Fuchs system should contain no first derivatives
with respect to $u_1$.

Note, that here $M_{km}$ is an arbitrary symmetric matrix.
By a special choice of $M_{km}$ we are able to diagonalize $L_{ij}$.
Really, there are $r(r+1)/2$ linear independent symmetric matrixes.
Adding the corresponding equations (\ref{picard0}) with unknown coefficients
we should set to zero $r(r-1)/2$ nondiagonal elements of $L_{ij}$. Thus we
obtain a system of $r(r-1)/2$ linear algebraic equations with $r(r+1)/2$
variables. It has $r=n-1$ linear independent solutions.

Therefore, (\ref{picard0}) can be rewritten as

\begin{equation}\label{picard1}
\left(\sum\limits_{k,m}M_{km}^{(p)}(u)
\frac{\partial^2}{\partial u_k \partial u_m}
+\sum\limits_{m\ne 1}K_m^{(p)}(u)\frac{\partial}{\partial u_m}
+ L^{(p)}(u)\right) \left(\begin{array}{c} a_i\\a_D^i\end{array}\right)= 0,
\end{equation}

\noindent
p=1,\ldots,n-1.

Let us check, that this system gives true and unique solution for

\begin{equation}
{\bf W}_k(u)\equiv a_D^i \frac{\partial a_i}{\partial u_k}
- a_i \frac{\partial a_D^i}{\partial u_k}.
\end{equation}

\noindent
It is easily verified, that for ${\bf W}$ (\ref{picard1}) gives

\begin{equation}\label{weq}
\sum\limits_{k,m} M_{km} \frac{\partial {\bf W}_k}{\partial u_m} +
\sum\limits_{m\ne 1} K_m {\bf W}_m = 0.
\end{equation}

Initial conditions are defined by the perturbative result, that is valid
if $u_1\to\infty$ \cite{seiberg,bilal}. Its form, of course, coincides
with (\ref{deriv1}):

\begin{equation}\label{der1}
{\bf W}_m(u_1\to\infty)=\frac{in}{\pi}\delta_{m1}.
\end{equation}

However, the relation between perturbative and nonperturbative anomalies
assumes, that (\ref{deriv1}) is satisfied for arbitrary $u_i$.
(\ref{picard1}) (and, therefore, in (\ref{weq})) should automatically
produce it.

Really, (\ref{der1}) is a solution of (\ref{weq}); there are
$n-1$ variables ${\bf W}_m$, which are uniquely determined from $n-1$
equations.

To conclude, the relation between perturbative and nonperturbative anomalies
in the N=2 supersymmetric SU(n) gauge theory leads to the system

\begin{equation}
\left(\sum\limits_{k,m}M_{km}^{(p)}(u)
\frac{\partial^2}{\partial u_k \partial u_m}
+\sum\limits_{m\ne 1}K_m^{(p)}(u)\frac{\partial}{\partial u_m}
+ L^{(p)}(u)\right) \left(\begin{array}{c} a_i\\a_D^i\end{array}\right)= 0,
\end{equation}

\noindent
p=1,\ldots,n-1.

These equations are in a complete agreement with the results of explicit
calculations (\ref{su2}) and (\ref{su3}) for SU(2) and SU(3) gauge groups.
(In particular, we explained the absence of first derivatives over
$u_1=u$, that seems accidental at the first sight.)


\subsection{Effective Lagrangian for N=1 supersymmetric theories.
General structure}\label{general_structure}
\hspace{\parindent}

Let us apply (\ref{relation}) to N=1 supersymmetric $SU(N_c)$ Yang-Mills
theory with $N_f$ matter supermultiplets. The vacuum expectation value
of the perturbative anomaly (\ref{pertanomaly}) is given by

\begin{eqnarray}\label{N1anomaly}
\langle \partial_\mu J^\mu_x\rangle
= \Big(N_f - N_c - x N_f\Big) \frac{1}{16\pi^2}
\mbox{Im}\int d^2\theta\ u.
\end{eqnarray}

\noindent
where $u\equiv \langle \mbox{tr} W^2\rangle$. Therefore, the effective
Lagrangian should depend in particular on $S= \mbox{tr} W^2$. This
result is not new. At the perturbative level the similar investigation
was made in \cite{veneziano}. However, in this paper we do not intend to
restrict ourselves by the frames of perturbation theory. Therefore, we
can not assume, that $u = \mbox{tr} W^2$ (Here we would like to remind
(\ref{correct})).

Comparing (\ref{N1anomaly}) with (\ref{N1anomaly1}) and taking into account,
that the equality should be satisfied for all $x$, we obtain

\begin{eqnarray}\label{2anomaly}
&&2w-2\frac{\partial w}{\partial S}S = - \frac{1}{16\pi^2}(N_f-N_c) u;
\nonumber\\
&&\frac{\partial w}{\partial v} v = - \frac{1}{16\pi^2} N_f u.
\end{eqnarray}

\noindent
This equation is very similar to the results of \cite{veneziano}.
Nevertheless, there is a crucial difference: $u\ne S$. Therefore,
one can only conclude that

\begin{equation}\label{ranom}
2w -2 \frac{\partial w}{\partial S} S
- \frac{N_f-N_c}{N_f}\frac{\partial w}{\partial v} = 0.
\end{equation}

This equation corresponds to the exact conservation of R-symmetry
at nonperturbative level. The similar condition was used in
\cite{affleck,seiberg2}, although the dependence $w=w(S)$ was ignored.
Of course, it is quite clear, that integrating out $S$ yields ADS
superpotential \cite{intriligator} and corresponds to imposing the
condition

\begin{equation}
\frac{\partial w}{\partial S} = 0.
\end{equation}

\noindent
We do not intend to discuss here the legitimacy of this operation and send
the reader to \cite{intriligator}, although we dare to suggest, that
the situation is more complicated. For the complete analysis we need
to know the kinetic part of the effective action, while now we can say
nothing about it, except for the general assumptions \cite{veneziano}. It
is worth to mention, that in the case $N_c > N_f$ the gauge group is not
completely broken and, therefore, the low energy theory contains
massless degrees of freedom at the perturbative level. We believe, that
integrating them out should be substantiated more thoroughly.

However this discussion, although being very interesting, is far
beyond the frames of the present paper. We would like only to stress,
that the presence of $S$-field is a strict consequence of (\ref{2anomaly}).

The solution of (\ref{ranom}) should agree with instanton calculations.
It is easily verified, that the only solution of (\ref{ranom}), agreeing
with (\ref{superpotential0})\footnote{It corresponds to
${\displaystyle g_n(x) = \frac{1}{32\pi^2} c_n x^{n(N_c-N_f)},
\quad n\ge 1}$
in (\ref{superpotential0}).}, is

\begin{equation}\label{solution}
w = \frac{1}{32\pi^2} S f(z);\qquad
f(z) = f_{pert}(z) + \sum\limits_{n=1}^\infty c_n z^n,
\end{equation}

\noindent
where

\begin{equation}
z\equiv\frac{\Lambda^{3N_c-N_f}}{v^{2N_f} S^{N_c-N_f}}
\end{equation}

\noindent
is a dimensionless parameter.

\noindent

In the final result $z$ should be written in terms of gauge invariant
variables. Of course, the result will depend on the structure of moduli
space, that is briefly reviewed in the Appendix \ref{moduli}. The
derivation, made in the Appendix \ref{derive_z}, gives

\begin{eqnarray}
&&\hspace{-7mm}
z=\frac{\Lambda^{3N_c-N_f}}{\mbox{det} M\ S^{N_c-N_f}}, \qquad N_f<N_c;
\nonumber\\
&&\hspace{-7mm}
z=\frac{\Lambda^{3N_c-N_f}S^{N_f-N_c}}{
\mbox{det} M - (\tilde B^{A_1 A_2\ldots A_{N_f-N_c}} M_{A_1}{}^{B_1}
M_{A_2}{}^{B_2} \ldots M_{A_{N_f-N_c}}{}^{B_{N_f-N_c}}
B_{B_1 B_2\ldots B_{N_f-N_c}})},\nonumber\\
&&\hspace{110mm}
N_f\ge N_c.
\end{eqnarray}

In order to define $f_{pert}$ we note, that at the perturbative level $u=S$.
Therefore, in this case (\ref{2anomaly}) gives

\begin{equation}\label{fpert}
\frac{\partial f_{pert}}{\partial z} z = 1,
\end{equation}

\noindent
so that

\begin{equation}\label{expansion}
w = \frac{1}{32\pi^2} S \Big(\ln z +
\sum\limits_{n=1}^\infty c_n z^n\Big).
\end{equation}

\noindent
Substituting it to (\ref{2anomaly}), we obtain

\begin{equation}\label{u_expansion}
u = S\Big(1 + \sum\limits_{n=1}^\infty n c_n z^n \Big),
\end{equation}

\noindent
that defines all anomalies in the theory according to (\ref{N1anomaly}).
At the perturbative level both (\ref{expansion}) and (\ref{u_expansion})
are certainly in agreement with \cite{veneziano}.

In the end of this section we should mention, that (\ref{expansion}) is not
in a complete agreement with instanton calculations. Really, although
$\Lambda^{(3N_c-N_f)n}/v^{2 N_f n}$ is already present in the instanton
measure, the result of integration over collective coordinates will
differ from the required form due to the factor exp(-$4\pi\rho^2 v^2$)
in the exponent. However the agreement can be achieved by changing the
form of the instanton vertex, but this problem is far beyond the frames
of present paper and we do not intend to discuss it here.


\subsection{Effective Lagrangian for N=1 supersymmetric theories.
Exact result}\label{exact_result}
\hspace{\parindent}

Let us try to define $f$ exactly. The general structure of the holomorphic
superpotential, found in section \ref{general_structure}, is similar to the
structure of the nonperturbative prepotential in the N=2 supersymmetric
Yang-Mills theory \cite{seiberg1}. In the latter case the relation between
perturbative and nonperturbative anomalies leads to Picard-Fuchs equations,
that can be used for derivation of exact results. Is it possible to extend
this approach to the case of N=1 supersymmetry?

First we substitute (\ref{solution}) into (\ref{2anomaly}), that gives

\begin{equation}\label{forf}
S \frac{df}{dz} z = u
\end{equation}

\noindent
(and therefore $u/S$ depends only on $z$).

The way to solve this equation is indicated by the analogy with N=2
supersymmetric $SU(2)$ Yang-Mills theory. In terms of N=1 superfields the
action (\ref{N2action}) is written as

\begin{equation}\label{N2versusN1}
\frac{1}{16\pi} \mbox{Im} \int d^4x d^2\theta \left(
\frac{d^2 F}{d\phi^2}W^2
+ \frac{1}{2}\int d^2\bar\theta \frac{dF}{d\phi}\phi^{+}\right).
\end{equation}

Let us compare it with

\begin{equation}\label{Sa}
S_a = \frac{1}{32\pi^2} \mbox{Re}\int d^4x d^2\theta\ S f(z)
\end{equation}

\noindent
and introduce $a\equiv z^{- 1/4}$ (this choice of the power will be
explained below). The first term in (\ref{N2versusN1}) will coincide
with (\ref{Sa}) if

\begin{equation}\label{new}
- 2\pi i \frac{d^2F}{da^2}\equiv f;
\qquad \frac{d^2U}{da^2}\equiv \frac{u}{S}.
\end{equation}

\noindent
Then (\ref{forf}) takes the form

\begin{equation}
F+F_D=\frac{2i}{\pi} U,
\end{equation}

\noindent
where

\begin{equation}
F_D=F-a a_D;\qquad a_D=\frac{dF}{da}.
\end{equation}

As above this leads to the Picard-Fuchs equation

\begin{equation}\label{pf}
\left(\frac{d^2}{da^2}+ L(U)\right)\left(\begin{array}{c}a\\a_D\end{array}
\right)=0,
\end{equation}

\noindent
where $L(U)$ is an undefined function.

At the perturbative level (see (\ref{fpert}))

\begin{equation}
\begin{array}{ll}
{\displaystyle f_{pert} = -4 \ln a\vphantom{\frac{1}{2}};}\\
{\displaystyle u_{pert} = W^2=S\vphantom{\frac{1}{2}},}
\end{array}
\qquad \mbox{so that}\qquad
\begin{array}{ll}
{\displaystyle F = \frac{i}{\pi} a^2
\Big(\frac{3}{2} - \ln a\Big);}\\
{\displaystyle U = a^2/2\vphantom{\frac{1}{2}}}
\end{array}
\end{equation}

\noindent
and, therefore

\begin{eqnarray}
&& a=\sqrt{2 U};\nonumber\\
&& a_D = - \frac{2i}{\pi} (a\ln a - a) =
- \frac{i}{\pi} \sqrt{2 U} \Big(\ln (2U) - 2\Big)
\end{eqnarray}

\noindent
are 2 independent solutions of the Picard-Fuchs equation

\begin{equation}
\left(\frac{d^2}{dU^2}+\frac{1}{4 U^2} \right)
\left(\begin{array}{c}a\\a_D\end{array}\right) = 0.
\end{equation}

However, the perturbative solution does not satisfy the requirement
\cite{seiberg,bilal}

\begin{equation}
\mbox{Im}\ \tau > 0,\qquad \mbox{where} \qquad
\tau = \frac{d^2F}{da^2} = \frac{da_D}{da} =\frac{i}{2\pi} f,
\end{equation}

\noindent
that is derived exactly as in the N=2 case. Therefore, two singularities
(at $U=0$ and $U=\infty$) are impossible.

To find the structure of singularities let us note, that the solution
(\ref{expansion}) should contain all positive powers of z and, therefore,
is invariant under $Z_4$ transformations $a \to e^{i\pi k/2} a$. Taking into
account (\ref{new}) and (\ref{u_expansion}) we conclude, that the
corresponding transformations in the $U$-plane are $U \to e^{i\pi k} U$.
Thus, singularities of $L(U)$ in the Picard-Fuchs equation (\ref{pf}) should
come in pairs: for each singularity at $U=U_0$ there is another one at
$U=-U_0$.

Therefore, the considered model is completely equivalent to N=2
supersymmetric SU(2) Yang-Mills theory without matter and the only
possible form of Picard-Fuchs equation (up to the redefinition of
$\Lambda$) is

\begin{equation}\label{N1pf}
\left(\frac{d^2}{dU^2}+\frac{1}{4 (U^2-1)} \right)
\left(\begin{array}{c}a\\a_D\end{array}\right) = 0
\end{equation}

\noindent
with the solution \cite{seiberg,bilal}

\begin{eqnarray}\label{final}
&&a(U)=\frac{\sqrt{2}}{\pi}\int\limits_{-1}^{1} dx
\frac{\sqrt{x-U}}{\sqrt{x^2-1}};
\nonumber\\
&&a_D(U)=\frac{\sqrt{2}}{\pi}\int\limits_{1}^{U} dx
\frac{\sqrt{x-U}}{\sqrt{x^2-1}}.
\end{eqnarray}

\noindent
Its uniqueness and, therefore, the uniqueness of the choice (\ref{N1pf})
was proven in \cite{unique}.

The function $F$ can be found by

\begin{equation}
\frac{dF}{dU} = a_D \frac{da}{dU}.
\end{equation}

\noindent
Its general structure is well known to be

\begin{equation}
F = - \frac{i}{\pi} a^2\Big(\ln a
+ \sum\limits_{n=0}^\infty F_n a^{-4n}\Big),
\end{equation}

\noindent
so that

\begin{equation}
f = -2\pi i \frac{d^2 F}{da^2} =
-4\ln a + \sum\limits_{n=0}^\infty f_n a^{-4n} =  \ln z +
\sum\limits_{n=0}^\infty f_n z^n.
\end{equation}

And now it is quite clear, that the choice $a = z^{-1/4}$ was made to
obtain the true structure of instanton corrections (\ref{solution}).


\section{Conclusion.}
\hspace{\parindent}

In the present paper we tried to investigate the structure of quantum
anomalies beyond the frames of perturbation theory. Although nontrivial
corrections exist due to the instanton effects, it is not difficult to
treat nonperturbative expressions. The matter is that the exact anomalies
turned out to be the vacuum expectation values of the perturbative ones.
However, the explicit check of this statement should essentially use the
structure of the result. Thus we are able to research nonperturbative effects
starting with this relation between perturbative and exact anomalies. The
approach was illustrated by deriving the Picard-Fuchs equations for the
exact solution of the SU(n) N=2 supersymmetric Yang-Mills theory. However,
the most interesting results were found when the method was applied
to N=1 theories. We managed to predict the general structure of holomorphic
superpotential and even to obtain the exact solution. Unfortunately, the
kinetic part can not be found by the presented approach. It complicates the
research significantly, because we are not able to solve some important
problems. In our opinion, the key question is when we can describe the
theory by the gauge invariant superfield $S$ and when it is necessary to
use original fields. It is really important, because the problem is tightly
bound with the quark confinement. We believe, that the solution can be
found only by the investigation of nonperturbative kinetic term, although
there are implications, that $S$ can be considered as a quantum field if
$N_f\ge N_c$ (see the brief discussion in the Section
\ref{general_structure}). Of course, it would be wonderful to solve this
problem and we do not lose the hope.

\vspace{1cm}

\noindent
{\Large\bf Acknowledgments}

\vspace{1cm}

The authors are very grateful to professor I.V.Tuitin and M.Alishahiha
for the valuable discussions and professors A.A.Slavnov, D.Bellisai and
M.Matone for the attention to the work. We especially like to thank
V.V.Asadov for the financial support.

\vspace{1cm}

\noindent
{\Large\bf Note added}

\vspace{1cm}

After sending this paper to hep-th we were informed by professor
M.Alishahiha and professor H.Schnitzer, that the general form of the
Picard-Fuchs equations for N=2 supersymmetric Yang-Mills theories was
already obtained in \cite{add1} by other methods. Our results agree with
it, so as with the similar results of \cite{add2}.


\vspace{1cm}

\noindent
{\Large\bf Appendix}

\appendix


\section{The classical moduli spaces of N=1 supersymmetric theories}
\label{moduli}
\hspace{\parindent}

To describe the vacuum states it is convenient to introduce two
$N_f\times N_c$ matrixes of the form

\begin{equation}
\phi \equiv \left(\phi^1, \phi^2,\ldots,\phi^{N_f}\right);\qquad
\tilde \phi \equiv \left(\tilde \phi_1, \tilde \phi_2,\ldots,
\tilde \phi_{N_f}\right).
\end{equation}

\noindent
(Their rows correspond to different values of color index.) The energy is
minimal if $\phi=\tilde \phi\equiv v$. Performing rotations in the color
and flavor spaces we can always reduce the matrix $v$ to the form

\begin{equation}
v=\left(
\begin{array}{cccc}
v_1    & 0      & \ldots & 0\\
0      & v_2    & \ldots & 0\\
\ldots & \ldots & \ldots & \ldots\\
0      & 0      & \ldots & v_{N_f}\\
0      & 0      & \ldots & 0\\
\ldots & \ldots & \ldots & \ldots
\end{array}
\right)
\end{equation}

\noindent
if $N_f < N_c$ and

\begin{equation}
v=\left(
\begin{array}{ccccccc}
v_1    & 0      & \ldots & 0       & 0      & \ldots \\
0      & v_2    & \ldots & 0       & 0      & \ldots \\
\ldots & \ldots & \ldots & \ldots  & \ldots & \ldots \\
0      & 0      & \ldots & v_{N_c} & 0      & \ldots
\end{array}
\right)
\end{equation}

\noindent
if $N_f > N_c$.

\vspace{5mm}

\noindent
1. $N_f < N_c$.

In the generic point the gauge group $SU(N_c)$ is broken down to
$SU(N_f-N_c)$. Therefore,

\begin{equation}
\Big(N_c^2-1\Big)-\Big((N_c-N_f)^2-1\Big) = 2 N_c N_f - N_f^2
\end{equation}

\noindent
chiral superfields are eaten up by super-Higgs mechanism. Taking into account
that originally there are $2 N_c N_f$ chiral matter superfields, we conclude
that only

\begin{equation}
2N_C N_f-\Big(2 N_c N_f - N_f^2\Big)=N_f^2
\end{equation}

\noindent
ones remain massless.

The flat direction can be described in the gauge invariant way by
$N_f^2$ composite chiral superfields

\begin{equation}
M_A{}^B = \tilde \phi_{Aa} \phi^{Ba}.
\end{equation}

\noindent
(Here $a$ denotes a color index.)

\vspace{5mm}

\noindent
2. $ N_f \ge N_c$.

If the number of flavors is equal to or larger than the number of colors,
the original gauge group is completely broken in the generic point.
Therefore, the number of remaining massless chiral superfields is

\begin{equation}
2 N_c N_f -\Big(N_c^2-1 \Big) = 2 N_c N_f - N_c^2 +1.
\end{equation}

In this case the gauge invariant description is provided by "mesons"

\begin{equation}
M_A{}^B = \tilde \phi_{Aa} \phi^{Ba}
\end{equation}

\noindent
and "barions"

\begin{eqnarray}
&&B_{A_{N_c+1}A_{N_c+2}\ldots A_{N_f}} = \frac{1}{N_c!}
\varepsilon_{A_1 A_2\ldots A_{N_f}} \varepsilon^{a_1 a_2\ldots a_{N_c}}
\phi^{A_1 a_1} \phi^{A_2 a_2} \ldots \phi^{A_{N_c} a_{N_c}};\nonumber\\
&&\tilde B^{A_{N_c+1}A_{N_c+2}\ldots A_{N_f}} = \frac{1}{N_c!}
\varepsilon^{A_1 A_2\ldots A_{N_f}} \varepsilon^{a_1 a_2\ldots a_{N_c}}
\phi_{A_1 a_1} \phi_{A_2 a_2} \ldots \phi_{A_{N_c} a_{N_c}}.
\end{eqnarray}

\noindent
However, their overall number is greater than $2 N_c N_f - N_c^2 +1$.
The matter is that at the classical level these fields are not
independent and satisfy some constraints. For example, if $N_f = N_c$
the number of massless superfields is $N_f^2+1$ while $N_M+N_B=N_f^2+2$.
The constraint eliminating the redundant chiral variable is

\begin{equation}
\mbox{det} M = \tilde B B.
\end{equation}

Similarly, for $N_f=N_c+1$

\begin{eqnarray}
&& B_A M_A{}^B = M_B{}^A \tilde B_A =0;\nonumber\\
&&\mbox{det} M \Big(M^{-1}\Big)_A{}^B = B_A \tilde B^B.
\end{eqnarray}

However, at the quantum level these constraints are violated by
instanton corrections and are no longer valid \cite{seiberg2}.


\section{On the gauge invariant form of parameter z}\label{derive_z}

\subsection{$N_f < N_c$}
\hspace{\parindent}

In this case the only gauge invariant parameter of $v^{2N_f}$ order
is det $M$, so that

\begin{equation}
z=\frac{\Lambda^{3N_c-N_f}}{\mbox{det} M\ S^{N_c-N_f}}.
\end{equation}

\subsection{$N_f \ge N_c$}
\hspace{\parindent}

For $N_f \ge N_c$ the moduli space is parametrized by mesons $M_A{}^B$ and
barions $B_{B_1\ldots B_{N_f-N_c}}$, $\tilde B^{A_1\ldots A_{N_f-N_c}}$,
satisfying some classical constrains. At the quantum level these constrains
are broken by instanton contributions. In the effective action approach the
modifications should be produced automatically. It can be achieved by
integrating out the $S$-superfield. \footnote{This implies, that $S$ becomes
a massive quantum field and the theory is described by the gauge invariant
variables or, by other words, confines.} (In particular, for $N_f=N_c$ $S$
is a natural Lagrange multiplier). The result should have the following form
\cite{peskin}:

\begin{eqnarray}\label{effective}
&& \mbox{det} M - \tilde B B = \mbox{const}\ \Lambda^{2 N_f},\qquad N_f=N_c;
\nonumber\\
\nonumber\\
&& w_{eff} = \mbox{const}\
\Lambda^{\textstyle - \frac{3N_c-N_f}{N_f-N_c}}
\Big(\mbox{det} M - (\tilde B^{A_1 A_2\ldots A_{N_f-N_c}} M_{A_1}{}^{B_1}
M_{A_2}{}^{B_2} \ldots M_{A_{N_f-N_c}}{}^{B_{N_f-N_c}}\nonumber\\
&&\times B_{B_1 B_2\ldots B_{N_f-N_c}})
\Big)^{\textstyle - \frac{1}{N_f-N_c}}+h.c.,
\qquad N_f > N_c.
\end{eqnarray}

\noindent
It can be achieved if and only if $v^{2N_f}$ is substituted by

\begin{equation}
\mbox{det} M - (\tilde B^{A_1 A_2\ldots A_{N_f-N_c}} M_{A_1}{}^{B_1}
M_{A_2}{}^{B_2} \ldots M_{A_{N_f-N_c}}{}^{B_{N_f-N_c}}
B_{B_1 B_2\ldots B_{N_f-N_c}}),
\end{equation}

\noindent
so that finally

\begin{equation}
z=\frac{\Lambda^{3N_c-N_f}S^{N_f-N_c}}{
\mbox{det} M - (\tilde B^{A_1 A_2\ldots A_{N_f-N_c}} M_{A_1}{}^{B_1}
M_{A_2}{}^{B_2} \ldots M_{A_{N_f-N_c}}{}^{B_{N_f-N_c}}
B_{B_1 B_2\ldots B_{N_f-N_c}})}.
\end{equation}

We would like to mention, that in the presented approach (\ref{effective})
certainly contains multiinstanton corrections, that contribute to the overall
constant factor in the RHS.



\begin{thebibliography}{200}

\bibitem{anomalybook}
R.Bertlmann, {\it Anomalies in quantum field theory}, Claredon Press,
Oxford, 1996.

\bibitem{matone}
M.Matone, {\it Phys.Lett.} {\bf B 357}, (1995), 342.

\bibitem{howe}
P.Howe, P.West, {\it Nucl.Phys.} {\bf B 486}, (1997), 425.

\bibitem{shifmanomaly}
V.Novikov, M.Shifman, A.Vainstein, V.Zakharov, {\it Nucl.Phys.}
{\bf B 229}, (1983), 394.

\bibitem{sonnenschein}
S.Sonnenschein, S.Theisen, S.Yankielowicz, {\it Phys.Lett.} {\bf B 367},
(1996), 145.

\bibitem{eguchi}
T.Eguchi, S.Yang, {\it Mod.Phys.Lett.}, {\bf A 11}, (1996), 145.

\bibitem{seiberg}
N.Seiberg and E.Witten, {\it Nucl.Phys.} {\bf B 426}, 19, (1994).

\bibitem{shifmanrev}
M.Shifman, {\it Prog.Part.Nucl.Phys.} {\bf 39}, 1, (1997).

\bibitem{klemm_review}
A.Klemm, {\it On the geometry behind N=2 supersymmetric effective actions
in four dimensions}, hep-th/9705131.

\bibitem{klemm1}
A.Klemm, W.Lerche, S.Yankielowicz, S.Theisen, {\it Phys.Lett.} {\bf B 344},
(1995), 169.

\bibitem{bilal}
A.Bilal, {\it Duality in N=2 SUSY SU(2) Yang-Mills theory: A pedagogical
introduction to the work of Seiberg and Witten}, hep-th/ 9601007.

\bibitem{klemm2}
A.Klemm, W.Lerche, S.Theisen, {\it Int.J.Mod.Phys.} {\bf A 11},
(1996), 1929.

\bibitem{thooft}
G.t'Hooft, {\it Phys.Rev.Lett.} {\bf 37}, (1976), 8;\\
G.t'Hooft, {\it Phys.Rev.} {\bf D 14}, (1976), 3432.

\bibitem{cordes}
S.Cordes, {\it Nucl.Phys.} {\bf B 273}, (1986), 629.

\bibitem{affleck}
I.Affleck, M.Dine, N.Seiberg, {\it Nucl.Phys.} {\bf B 241}, (1984), 493.

\bibitem{shifman}
M.Shifman, A.Vainstein, V.Zakharov, {\it Nucl.Phys.} {\bf B 165}, (1980),
45;\\
V.Novikov, M.Shifman, A.Vainstein, V.Zakharov, {\it Nucl.Phys.} {\bf B 229},
(1983), 407;\\
V.Novikov, M.Shifman, A.Vainstein, V.Zakharov, {\it Nucl.Phys.} {\bf B 260},
(1985), 157.

\bibitem{yung}
A.Yung, {\it Nucl.Phys.} {\bf B 485}, 38, (1997).

\bibitem{callan}
C.Callan, R.Dashen, D.Gross, {\it Phys.Rev.} {\bf D 17}, (1978), 2717.

\bibitem{veneziano}
G.Veneziano, S.Yankelowicz, {\it Phys.Let.} {\bf B 113}, (1982), 231;\\
T.Taylor, G.Veneziano, S.Yankelowicz, {\it Nucl.Phys.} {\bf B 218}, (1983),
493.

\bibitem{bellisai}
D. Bellisai, F. Fucito, M. Matone, G. Travaglini, {\it Phys. Rev.}
{\bf D 56}, (1997), 5218.

\bibitem{finnell}
D.Finnell, P.Pouliot, {\it Nucl.Phys.} {\bf B 453}, 225, (1995).

\bibitem{fujikava}
K.Fujikava, {\it Phys.Rev.Lett.} {\bf 42}, 1195, (1979);\\
K.Fujikava, {\it Phys.Rev.Lett.} {\bf 44}, 1733, (1980).

\bibitem{seiberg2}
N.Seiberg, {\it Phys.Rev.} {\bf D 49}, (1994), 6857.

\bibitem{intriligator}
K.Intriligator, R.Leith, N.Seiberg, {\it Phys.Rev.} {\bf D 50}, (1994), 1092.

\bibitem{seiberg1}
N.Seiberg, {\it Phys.Lett.} {\bf B 206}, 75, (1988).

\bibitem{unique}
G. Bonelli, M.Matone, M. Tonin, {\it Phys. Rev.} {\bf D 55}, (1997), 6466;\\
R.Flume, M.Magro, L.O'Raifeartaigh, I.Sachs, O.Schnetz, {\it Nucl.Phys.}
{\bf B 494}, (1997), 331.

\bibitem{peskin}
M.Peskin, {\it Duality in supersymmetric Yang-Mills theory}, hep-th/9702094.


\bibitem{add1}
M.Alishahiha, {\it Simple derivation of the Picard-Fuchs equations for
the Seiberg-Witten models}, hep-th/9703186;

J.Isidro, A.Mukherjee, J.Nunes, H.Schnitzer, {\it Int.J.Mod.Phys.}
{\bf A 13}, (1998), 233.

\bibitem{add2}
M.Alishahiha, {\it Phys.Lett.} {\bf B 398}, 100;

J.Isidro, A.Mukherjee, J.Nunes, H.Schnitzer, {\it Nucl.Phys.} {\bf B 492},
(1997), 647.

\end{thebibliography}
\end{document}